# Hierarchical Empirical-Bayes Forecasting of Sparse Counts with Exposure:
# An Application to Airline Demand and Premium-Cabin Capacity Sharing

Irakli Gudavadze

October 5, 2026

**Abstract**

Sparse count forecasting arises when events are rare, exposure varies across observational units, and short histories contain many zeros. Using thin-market airline demand as a motivating application, this paper develops a hierarchical exposure-adjusted Poisson-Gamma (HEPG) framework for probabilistic forecasting. Conditional counts follow a Poisson model with an exposure offset and pooled calendar effects, while unit-specific baseline intensities receive a Gamma prior whose hyperparameters are estimated empirically across the population. The resulting posterior is Gamma and the posterior predictive distribution is negative binomial, providing closed-form partial pooling and explicit propagation of parameter uncertainty. This is especially useful for zero-history or low-exposure units, where independent maximum-likelihood estimates are unstable or collapse to zero. An adaptive extension, A-HEPG, adds a moment-matched negative-binomial variance correction when residual extra-Poisson variation remains. In synthetic experiments with 400 heterogeneous markets, HEPG improves average negative log predictive probability relative to complete pooling and independent route-level Poisson estimation. At a mean latent rate of 0.15 requests per exposure unit, the score is 0.525 for HEPG, compared with 0.585 for pooled Poisson and 0.731 for route-level Poisson maximum likelihood. Under strong overdispersion, A-HEPG reduces the score from 0.552 to 0.529. A downstream cabin-capacity example links the predictive distribution to a seat-protection rule based on predictive survival probabilities and fare ratios. The study emphasizes coherent probabilistic inference, partial pooling, and diagnostic model escalation for sparse applied count data.

## Introduction

Sparse count forecasting is a recurring applied-statistics problem whenever events are rare relative to the number of observation opportunities. Examples include failures, claims, disease events, intermittent purchases, and low-frequency booking requests. In airline revenue management, the same problem appears at fine levels of granularity such as an origin-destination market, flight leg, fare family, booking-window interval, or newly launched service. At high volumes, continuous approximations can be adequate. At low volumes, however, the statistical structure changes qualitatively. A short history such as

$$0,\ 0,\ 1,\ 0,\ 0,\ 0,\ 2,\ 0,\ 0,\ldots$$

is not well represented by a symmetric continuous distribution. Zero is a valid and often probable outcome, the support is nonnegative and integer-valued, and sampling uncertainty can be large relative to the mean. These features make both the likelihood and the treatment of parameter uncertainty important.

The airline setting gives these statistical issues a direct decision consequence. Forecasts feed seat protection, availability, pricing, schedule evaluation, and other revenue-management decisions (Talluri & Ryzin, 2004; Weatherford & Belobaba, 2002). A point estimate alone is often insufficient: the probability of zero demand, the probability of at least one request, and upper predictive quantiles can all influence decisions. Thin markets also create a small-sample estimation problem. Estimating each market independently produces noisy rates and can assign a rate of exactly zero after a zero-count training history, whereas complete pooling avoids zero estimates at the cost of erasing genuine cross-market heterogeneity.

The natural statistical starting point is a count model. Poisson processes are established tools in both airline revenue management and the wider intermittent-demand literature (Hung & Chen, 2013; Weatherford & Pölt, 2002; Snyder et al., 2012). The contribution of this paper is therefore not a claim that the Poisson distribution, Poisson-Gamma conjugacy, or empirical Bayes is new. Rather, the paper develops an application-focused sparse-count workflow that combines five elements:

1. an exposure-aware count likelihood, allowing observational units with different numbers of selling opportunities or scheduled departures to be compared on a common rate scale;
2. empirical-Bayes hierarchical partial pooling across related units, stabilizing sparse estimates without forcing all units to share one rate;
3. a closed-form posterior predictive distribution that propagates rate uncertainty and automatically widens when unit-level information is scarce;
4. a diagnostic variance correction that relaxes conditional equidispersion when residual extra-Poisson variation remains; and
5. an applied decision layer showing how a sparse predictive distribution can be converted into a transparent premium-cabin seat-protection rule based on predictive tail probabilities and fare differences.

The framework is called the *hierarchical exposure-adjusted Poisson–Gamma* model, abbreviated HEPG. If market $i$ has latent rate $\lambda_i$, period $t$ has exposure $e_{it}$, and pooled calendar multiplier $s_{it}$, then

$$Y_{it} \mid \lambda_i \sim \text{Poisson}(e_{it} s_{it} \lambda_i), \qquad \lambda_i \sim \text{Gamma}(a, b),$$

where the Gamma distribution is parameterized by shape $a$ and rate $b$. Conditional on the pooled covariate effects, this hierarchy is conjugate. The posterior rate estimate is a weighted average of the route-only estimate and the network prior mean. For a future period, integrating over route uncertainty yields a negative-binomial predictive distribution even though the conditional arrival mechanism is Poisson.

This structure is particularly useful in the low-count regime because parameter uncertainty is a first-order component of predictive uncertainty. When a unit has only a few effective exposure units, a forecast should be wider than a plug-in Poisson distribution based on a single estimated mean. HEPG obtains that widening analytically through the posterior predictive distribution rather than through an ad hoc safety factor.

The application also highlights a data-quality distinction that is broadly relevant to applied count modeling: an observed zero is not always a realization of zero latent demand. In airline data, a zero can occur because a flight did not operate, inventory was closed, a fare product was unavailable, or bookings were censored by controls. The model is intended for latent request counts, open and uncensored observations, or bookings that have been appropriately unconstrained. Treating an availability-induced zero as a stochastic demand zero biases any count model.

The remainder of the paper reviews the statistical and application literature, formulates the HEPG and adaptive HEPG models, studies their predictive behavior in controlled simulations, illustrates a downstream decision rule, and discusses validation and deployment issues for observational airline data.

## Related statistical and application literature

### Airline demand forecasting as an applied count problem

Revenue management converts uncertain demand forecasts into inventory and pricing decisions. Talluri & Ryzin (2004) provide a broad treatment of the field, while Weatherford & Belobaba (2002) show that demand forecast error can materially affect the performance of airline seat-allocation rules. Forecasting is complicated by censoring: historical bookings can be lower than latent demand because customers are denied availability or because capacity is exhausted. Weatherford & Pölt (2002) and Queenan et al. (2007) analyze methods for recovering unconstrained demand from such observations.

Distributional assumptions have also received explicit attention. Swan (2002) discusses airline demand distributions and argues that the shape of demand depends on the underlying arrival and heterogeneity mechanisms. Poisson arrival assumptions have appeared in airline revenue-management models, including nonhomogeneous Poisson customer arrivals (Hung & Chen, 2013). More recent work has combined unconstraining and machine-learning forecasts for airline revenue management (Wang et al., 2025). These contributions largely address general airline forecasting or control; the present paper focuses specifically on the sparse-count regime where route-level estimation uncertainty and zero outcomes dominate.

A closely related operational problem is premium-cabin capacity sharing. Escovar-Álvarez & Belobaba (2022) study the acceptance of additional economy bookings when premium seats are expected to remain empty and report simulated revenue gains, while emphasizing the risk of displacing later premium-fare demand and the importance of applying the strategy late in the booking horizon. Feng et al. (2024) analyze dynamic capacity sharing and an extension in which economy passengers can receive free upgrades into business-class capacity. These studies establish that cabin sharing is a genuine revenue-management problem. The contribution here is narrower: we show how a sparse premium-demand predictive distribution can feed an analytically transparent protection rule whose aggressiveness changes with the premium–economy fare gap.

## Intermittent-demand and sparse-count forecasting

Sparse airline booking series are related to intermittent-demand problems in inventory forecasting. Croston (1972) proposed separate smoothing of demand size and inter-arrival time. Subsequent work studied the bias and stochastic interpretation of Croston-type procedures (Shenstone & Hyndman, 2005; Syntetos & Boylan, 2005). For probabilistic planning, Snyder et al. (2012) compare Poisson, negative-binomial, and hurdle-type models and emphasize the value of modeling the full predictive distribution rather than only a point forecast.

The count literature also provides extensions for excess zeros and overdispersion. Zero-inflated count time-series models have been developed for settings where zeros occur more often than a standard Poisson model predicts (Maiti et al., 2015). Jiang (2020) separates zero and positive-demand mechanisms using a mixed zero-truncated Poisson approach. More recently, Aktekin et al. (2026) developed Bayesian zero-inflated Poisson–Gamma state-space models for sequential count forecasting. These models motivate a diagnostic principle used in this paper: begin with the simplest coherent count process, then relax it only when residual diagnostics show that Poisson variability is insufficient.

## Hierarchical Poisson and empirical-Bayes modeling

Hierarchical Poisson models and empirical-Bayes estimation are classical statistical tools for count data (Cameron & Trivedi, 2013; Tu & Piegorsch, 2003). Their relevance to sparse applications is direct: a population contains related but nonidentical units, and the analyst must trade off complete pooling against unstable unit-by-unit estimation. A hierarchy lies between these extremes, allowing low-information units to borrow more strength while letting mature units depend primarily on their own histories.

The statistical ingredients are established rather than theoretically novel. The applied contribution is their assembly into an exposure-aware sparse-count forecasting workflow with explicit treatment of zero-history units, posterior predictive uncertainty, proper probabilistic scoring, residual overdispersion diagnostics, and the distinction between stochastic zeros and zeros created by availability or censoring. These issues are central in the airline application and arise in many other sparse-count settings.

## Statistical formulation

Let $i = 1, \dots, I$ index forecasting cells. A cell can be a route, leg, fare product, point-of-sale segment, or another stable market definition. Let $t = 1, \dots, T$ index time periods. The latent variable of interest is

$$Y_{it} = \text{number of demand requests in cell } i \text{ during period } t.$$

The period should be chosen so that low demand is meaningful at the decision granularity. Examples include daily requests for a thin market, requests in a days-to-departure bucket, or bookings for a narrow fare product. We call a cell *thin* when its expected count per modeling interval is small enough that zeros and single-request outcomes have material probability. No universal numerical threshold is required, but expected counts below one are a natural regime in which discrete modeling becomes especially important.

### Exposure

Let $e_{it} \geq 0$ denote known exposure. Depending on the application, exposure can represent the number of bookable departures, the duration of an observation window, or another quantity proportional to opportunity for requests. If every observation has the same opportunity, set $e_{it} = 1$.

Exposure is not a substitute for availability logic. If no service is offered, $e_{it} = 0$ and the record should not be interpreted as evidence of zero latent demand at a positive exposure level. Likewise, if a booking class is closed, the analyst must distinguish "no observed bookings" from "no customer requests."

### Calendar and demand multipliers

Let $x_{it}$ contain known predictors such as day-of-week indicators, seasonal harmonics, holiday flags, booking-horizon effects, or market-level covariates. Define

$$s_{it}(\beta) = \exp(x_{it}^{\top}\beta).$$

The conditional mean is

$$\mu_{it} = e_{it} s_{it}(\beta) \lambda_i.$$

This formulation separates a market-specific baseline intensity $\lambda_i$ from multiplicative effects shared across many markets. Richer specifications can include interactions or hierarchical coefficients, but the simple decomposition preserves interpretability and conjugacy in $\lambda_i$ conditional on $\beta$.

## Hierarchical exposure-adjusted Poisson–Gamma model

### Conditional count model

The core observation model is

$$Y_{it} \mid \lambda_i, \beta \sim \text{Poisson}(e_{it} s_{it}(\beta) \lambda_i).$$

For genuinely low counts, the Poisson model has three attractive properties. It has the correct support, it assigns positive probability to zero, and its likelihood naturally weights information by exposure. Its main restriction is equidispersion conditional on the latent rate:

$$\text{Var}(Y_{it} \mid \lambda_i, \beta) = \text{E}(Y_{it} \mid \lambda_i, \beta).$$

The adaptive extension below relaxes this restriction.

## Network prior and partial pooling

Market intensities are modeled as

$$\lambda_i \sim \text{Gamma}(a, b),$$

with shape $a$ and rate $b$. Thus

$$\text{E}(\lambda_i) = \frac{a}{b}, \qquad \text{Var}(\lambda_i) = \frac{a}{b^2}.$$

For fixed $\beta$, define the total count and effective exposure

$$Y_i = \sum_{t=1}^{T} Y_{it}, \qquad S_i = \sum_{t=1}^{T} e_{it}\, s_{it}(\beta).$$

Conjugacy gives

$$\lambda_i \mid \mathcal{D} \sim \text{Gamma}(A_i, B_i), \qquad A_i = a + Y_i, \quad B_i = b + S_i.$$

The posterior mean has an illuminating shrinkage form:

$$\hat{\lambda}_{i,\text{HEPG}} = \frac{a + Y_i}{b + S_i}.$$

Equivalently,

$$\hat{\lambda}_{i,\text{HEPG}} = w_i \frac{Y_i}{S_i} + (1 - w_i)\frac{a}{b}, \qquad w_i = \frac{S_i}{b + S_i}.$$

The weight $w_i$ depends only on effective exposure. A mature route with large $S_i$ is barely shrunk. A sparse route with small $S_i$ is pulled toward the network distribution. Most importantly, if $Y_i = 0$, HEPG does not force the future rate to zero. The forecast remains positive unless the learned network prior itself degenerates at zero.

The shrinkage figure below illustrates this behavior.

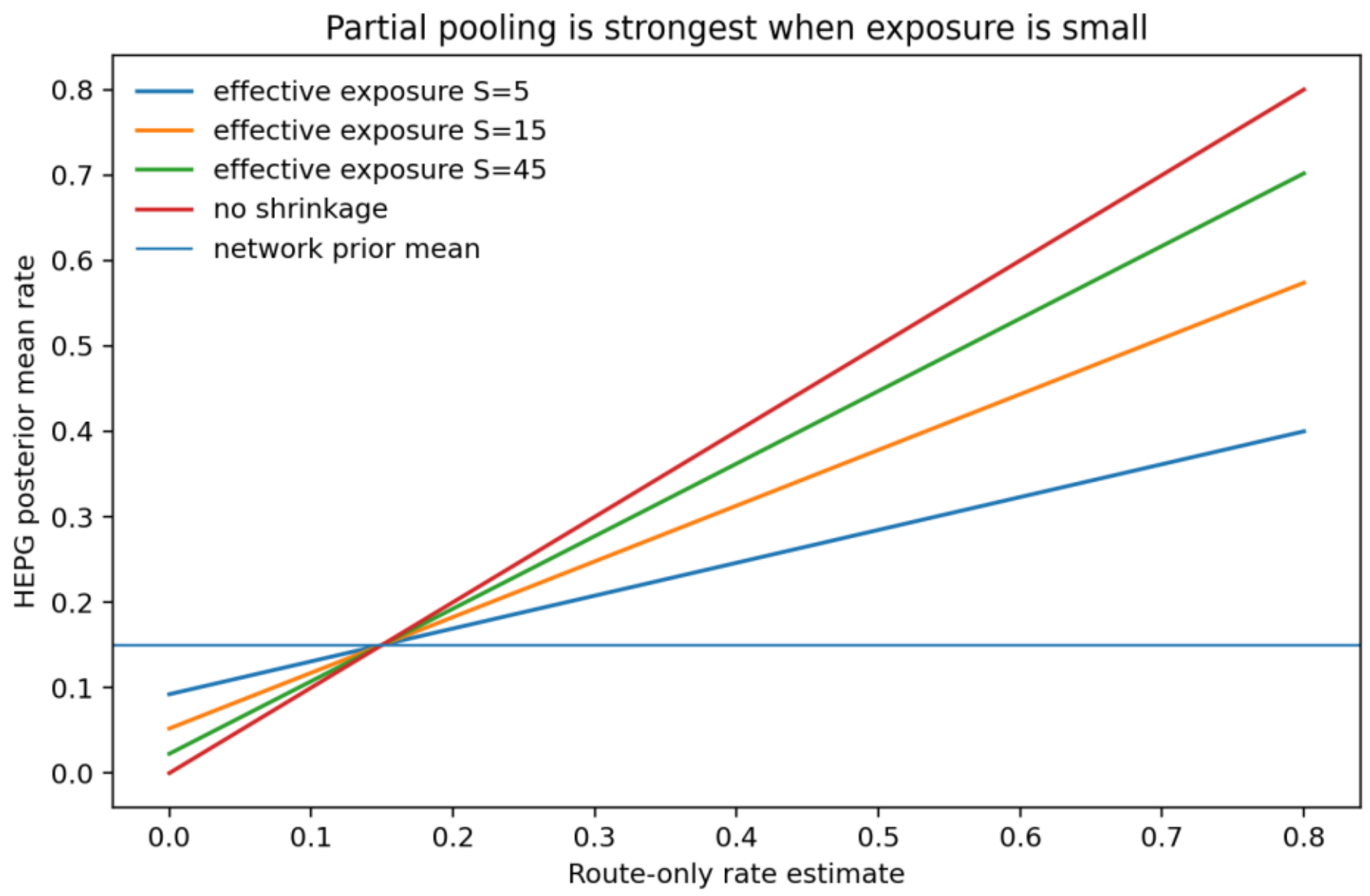

*HEPG posterior mean as a function of the route-only estimate. Partial pooling is strongest when effective exposure is small and vanishes as route information accumulates.*

## Posterior predictive distribution

For a future period with

$$c_{i,f} = e_{i,f} s_{i,f}(\beta),$$

integrating the Poisson likelihood over the Gamma posterior gives

$$Y_{i,f} \mid \mathcal{D} \sim \mathrm{NegBin}\left(A_i,\ p_i = \frac{B_i}{B_i + c_{i,f}}\right),$$

using the convention in which the negative-binomial random variable counts events before $A_i$ successes.

The predictive mean and variance are

$$m_{i,f} = c_{i,f} \frac{A_i}{B_i},$$

and

$$v_{i,f} = m_{i,f} + \frac{m_{i,f}^2}{A_i}.$$

The second term in the variance is parameter uncertainty. When $A_i$ is small, uncertainty about the sparse route rate materially widens the predictive distribution. As the route accumulates history, $A_i$ grows and the predictive variance approaches the plug-in Poisson variance.

This is a central practical advantage of the hierarchy. It retains a Poisson arrival interpretation but does not pretend that an imprecisely estimated rate is known exactly.

### Empirical-Bayes estimation

The hyperparameters $(a, b)$ are learned from the network rather than fixed arbitrarily. Conditional on $\beta$, integrating out $\lambda_i$ gives the route-level marginal likelihood

$$p(\mathbf{y}_i \mid a, b, \beta) = \frac{b^a \Gamma(a + Y_i)}{\Gamma(a)(b + S_i)^{a+Y_i}} \prod_{t=1}^{T} \frac{[e_{it} s_{it}(\beta)]^{Y_{it}}}{Y_{it}!}.$$

The empirical-Bayes estimator maximizes the product of the marginal likelihood above over markets.

Two implementation strategies are useful. First, one may jointly optimize $(a, b, \beta)$. Second, for operational simplicity, one may estimate pooled calendar effects first, compute $S_i$, and then estimate $(a, b)$ by marginal likelihood. The simulation study uses the second strategy. In production, joint optimization or alternating updates are preferable when exposure and route intensity are correlated.

## Robustness to extra-Poisson variation

A Poisson model can be too narrow when demand contains multiplicative shocks, group bookings, unobserved event effects, or other sources of heterogeneity. The usual negative-binomial variance function is

$$\mathrm{Var}(Y \mid \mu) = \mu + \frac{\mu^2}{k},$$

where smaller $k$ means stronger overdispersion.

We use the HEPG posterior mean as the stable route-intensity estimate and diagnose remaining extra-Poisson variation with the moment estimator

$$\hat{\theta} = \max\left\{0,\ \frac{\sum_{it} [(Y_{it} - \hat{\mu}_{it})^2 - \hat{\mu}_{it}]}{\sum_{it} \hat{\mu}_{it}^2}\right\}, \qquad \hat{k} = \hat{\theta}^{-1}.$$

When $\hat{\theta}$ is essentially zero, $\hat{k}$ is treated as very large and the model reduces to HEPG.

For a variance-robust predictive approximation, suppose a multiplicative shock $U$ has

$$U \sim \mathrm{Gamma}(k, k), \qquad E(U) = 1, \quad \mathrm{Var}(U) = 1/k.$$

Combining $U$ with posterior route uncertainty yields a predictive mean $m_{i,f}$ and an overdispersed variance. A negative-binomial distribution can be moment-matched to these two moments using effective size

$$r_{i,f} = \frac{A_i k}{k + A_i + 1}.$$

We call this predictive correction *adaptive HEPG* (A-HEPG). It should be interpreted as a robust moment-matched predictive approximation, not as an exact conjugate posterior for a negative-binomial observation model.

## When zero inflation or hurdle models are preferable

The Poisson distribution naturally generates zeros. Therefore a large number of zeros is not, by itself, evidence that a zero-inflated model is needed. The appropriate diagnostic is whether the observed zero frequency materially exceeds the zero probability implied by a well-fitted count model after known no-service and unavailable-inventory periods have been removed.

If excess zeros remain, a two-process model may be more appropriate:

$$P(Y_{it} = 0) = \pi_{it} + (1 - \pi_{it}) f_{\text{count}}(0),$$

where $\pi_{it}$ is a structural-zero probability. A hurdle alternative models $P(Y_{it} > 0)$ separately and uses a zero-truncated count distribution for positive outcomes. These ideas are well developed in the intermittent-demand literature (Aktekin et al., 2026; Jiang, 2020; Maiti et al., 2015).

For airline applications, however, "structural zero" must be interpreted carefully. A canceled or nonexistent service should typically be represented as zero exposure, not as a stochastic structural-demand state. Likewise, a fare product that is closed by inventory control produces censored observation, not necessarily structural zero demand.

## Censoring and unconstraining

The core model assumes that $Y_{it}$ represents latent requests or an observation that equals latent demand. When accepted bookings are constrained, the likelihood must reflect censoring or the data must first be unconstrained.

For a simple right-censoring threshold $C_{it}$, the contribution to the likelihood is

$$L_{it} = \begin{cases} P(Y_{it} = y_{it} \mid \mu_{it}), & \text{if uncensored,} \\ P(Y_{it} \geq C_{it} \mid \mu_{it}), & \text{if censored.} \end{cases}$$

The same principle applies to negative-binomial or hurdle extensions. In a production airline setting, demand preprocessing and availability reconstruction can be as important as the choice of forecast distribution. The simulation study below intentionally removes this complication to isolate the sparse-count estimation problem.

## Simulation study

### Objectives

The simulation has three statistical goals. First, it tests whether hierarchical partial pooling improves probabilistic forecasts when the underlying units are heterogeneous but individually sparse. Second, it examines the zero-history failure mode of independent maximum-likelihood estimation. Third, it evaluates whether a simple adaptive variance correction improves calibration and predictive score when the conditional Poisson assumption is violated.

All simulations are synthetic. No proprietary airline data are used.

### Data-generating process

Each replication contains $I = 400$ markets. Baseline route intensities are drawn from

$$\lambda_i \sim \text{Gamma}\left(1.2, \frac{1.2}{\bar{\lambda}}\right),$$

so that $E(\lambda_i) = \bar{\lambda}$. The Poisson experiments use

$$\bar{\lambda} \in \{0.08, 0.15, 0.30\}$$

requests per exposure unit.

Daily exposure starts from a route-level frequency of one, two, or three units with probabilities 0.60, 0.30, and 0.10, respectively, plus small random day-to-day variation. Weekly seasonality follows

$$s_t \propto \exp[0.25 \sin(2\pi t/7) - 0.10 \cos(2\pi t/7)],$$

normalized to have weekly mean one.

For the correctly specified experiments,

$$Y_{it} \sim \text{Poisson}(e_{it} s_t \lambda_i).$$

Each replication uses 14 training days and 28 test days, deliberately creating a short-history environment.

The robustness experiments use negative-binomial observations with mean $e_{it} s_t \lambda_i$ and size

$$k \in \{5,\ 2,\ 0.8\},$$

with smaller $k$ representing stronger extra-Poisson variation. These experiments use 28 training days and 28 test days.

Thirty independent replications are run for every scenario. Pooled weekly seasonality is estimated from the training sample in each replication.

## Competing methods

Five forecasting methods are compared:

1. **Pooled Poisson**: one network-wide rate shared by all markets.
2. **Route Poisson MLE**: a separate maximum-likelihood rate for each route. For numerical scoring, zero estimates are clipped at a negligible positive value.
3. **Weak-prior Poisson–Gamma**: independent route models with a fixed Gamma(0.5,0.5) prior, illustrating generic smoothing without network-learned hyperparameters.
4. **HEPG**: the proposed empirical-Bayes Poisson–Gamma model.
5. **A-HEPG**: HEPG with the extra-Poisson variance correction described above.

## Evaluation metrics

The primary metric is average negative log predictive probability,

$$\mathrm{NLL} = -\frac{1}{N_{\mathrm{test}}}\sum \log p(Y_{it}^{\mathrm{test}} \mid \mathcal{D}),$$

a proper scoring rule for the full predictive distribution. Lower is better.

We also report mean absolute error (MAE), the empirical coverage of central 90% predictive intervals, and the Brier score for the zero event. Because very low-count distributions are discrete, nominal equal-tailed intervals are expected to be conservative; coverage should therefore be interpreted together with interval width and proper scores.

# Results

## Sparse Poisson demand

The table below reports results under the correctly specified Poisson data-generating process.

*Synthetic Poisson experiments. Lower NLL, MAE, and Brier scores are better. Coverage is for central 90% predictive intervals.*

| Mean rate | Method | NLL | MAE | Coverage | Zero Brier |
|---|---|---|---|---|---|
| 0.08 | Pooled Poisson | 0.381 | 0.215 | 0.987 | 0.095 |
| | Route Poisson MLE | 0.608 | **0.193** | 0.975 | 0.092 |
| | Weak-prior PG | 0.362 | 0.218 | 0.979 | 0.091 |
| | HEPG | **0.355** | 0.201 | 0.981 | **0.090** |
| | A-HEPG | **0.355** | 0.201 | 0.981 | **0.090** |

| Mean rate | Method | NLL | MAE | Coverage | Zero Brier |
|---|---|---|---|---|---|
| 0.15 | Pooled Poisson | 0.585 | 0.362 | 0.973 | 0.146 |
| | Route Poisson MLE | 0.731 | **0.304** | 0.972 | 0.133 |
| | Weak-prior PG | 0.530 | 0.324 | 0.977 | 0.132 |
| | HEPG | **0.525** | 0.316 | 0.981 | **0.131** |
| | A-HEPG | **0.525** | 0.316 | 0.981 | **0.131** |
| 0.30 | Pooled Poisson | 0.909 | 0.597 | 0.955 | 0.207 |
| | Route Poisson MLE | 0.899 | **0.470** | 0.970 | 0.170 |
| | Weak-prior PG | 0.774 | 0.483 | 0.975 | 0.169 |
| | HEPG | **0.770** | 0.484 | 0.978 | **0.169** |
| | A-HEPG | **0.770** | 0.484 | 0.978 | **0.169** |

At $\bar{\lambda} = 0.15$, HEPG reduces average NLL from 0.585 for pooled Poisson to 0.525, a reduction of about 10%. Relative to route-level Poisson MLE, the reduction is about 28%. At $\bar{\lambda} = 0.30$, the HEPG NLL is 0.770 versus 0.909 for pooled Poisson and 0.899 for route-level MLE.

The route-level MLE has slightly lower MAE in these experiments. This is an important limitation and not a contradiction. The route MLE is a sharp point estimate and, under absolute error, can occasionally outperform a posterior mean even while assigning implausibly small probability to unseen positive counts after a zero history. NLL penalizes such overconfidence. The principal HEPG benefit is therefore probabilistic rather than universal point-error dominance.

The weak-prior Poisson–Gamma model already removes the catastrophic zero-rate problem. HEPG improves modestly over that baseline because it learns the amount and center of shrinkage from the network instead of fixing the prior arbitrarily.

The following figure shows how the probabilistic advantage changes with the average rate.

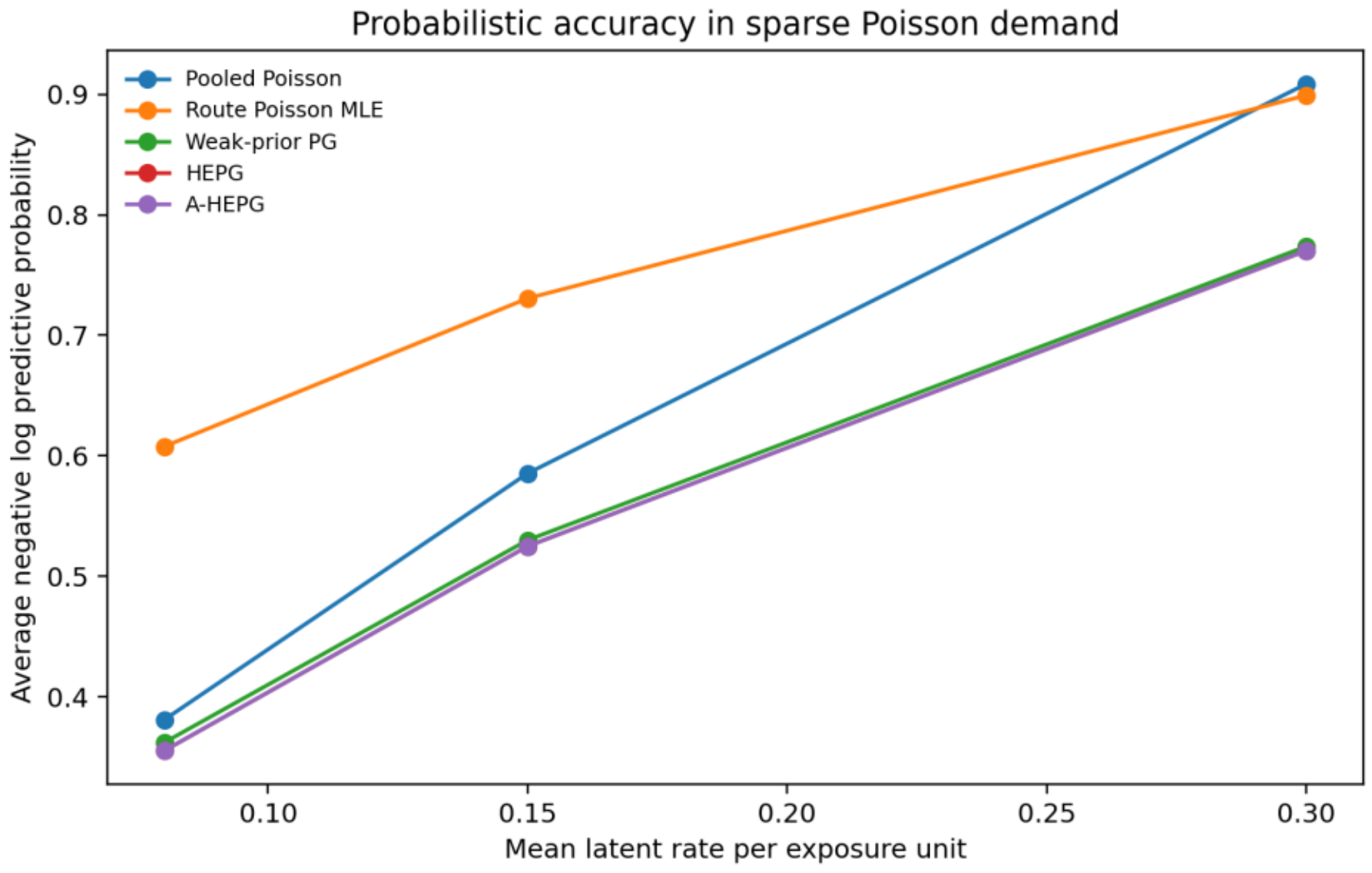

*Average negative log predictive probability in sparse Poisson simulations. Lower is better.*

## Robustness to overdispersion

The table below reports the negative-binomial robustness experiments.

*Synthetic overdispersed experiments at mean rate 0.15. Smaller true k indicates stronger overdispersion.*

| True $k$ | Method | NLL | MAE | Coverage | Zero Brier |
|---|---|---|---|---|---|
| 5.0 | Pooled Poisson | 0.590 | 0.366 | 0.972 | 0.144 |
| | Route Poisson MLE | 0.586 | **0.311** | 0.973 | 0.129 |
| | Weak-prior PG | 0.524 | 0.321 | 0.978 | 0.128 |
| | HEPG | 0.521 | 0.318 | 0.978 | 0.128 |
| | A-HEPG | **0.520** | 0.318 | 0.979 | **0.128** |
| 2.0 | Pooled Poisson | 0.597 | 0.369 | 0.969 | 0.140 |
| | Route Poisson MLE | 0.598 | **0.318** | 0.970 | 0.127 |
| | Weak-prior PG | 0.532 | 0.329 | 0.975 | 0.127 |
| | HEPG | 0.529 | 0.325 | 0.975 | 0.126 |
| | A-HEPG | **0.524** | 0.325 | 0.978 | **0.126** |
| 0.8 | Pooled Poisson | 0.624 | 0.381 | 0.964 | 0.134 |
| | Route Poisson MLE | 0.627 | **0.337** | 0.963 | 0.125 |
| | Weak-prior PG | 0.554 | 0.348 | 0.968 | 0.125 |
| | HEPG | 0.552 | 0.342 | 0.968 | 0.124 |
| | A-HEPG | **0.529** | 0.342 | 0.974 | **0.121** |

When overdispersion is mild ($k = 5$), A-HEPG and HEPG are nearly identical. At $k = 2$, A-HEPG lowers NLL from 0.529 to 0.524. Under strong overdispersion ($k = 0.8$), the improvement is larger: NLL falls from 0.552 to 0.529. This pattern is desirable. The adaptive correction changes little when the Poisson approximation is adequate and matters more when the variance materially exceeds the mean.

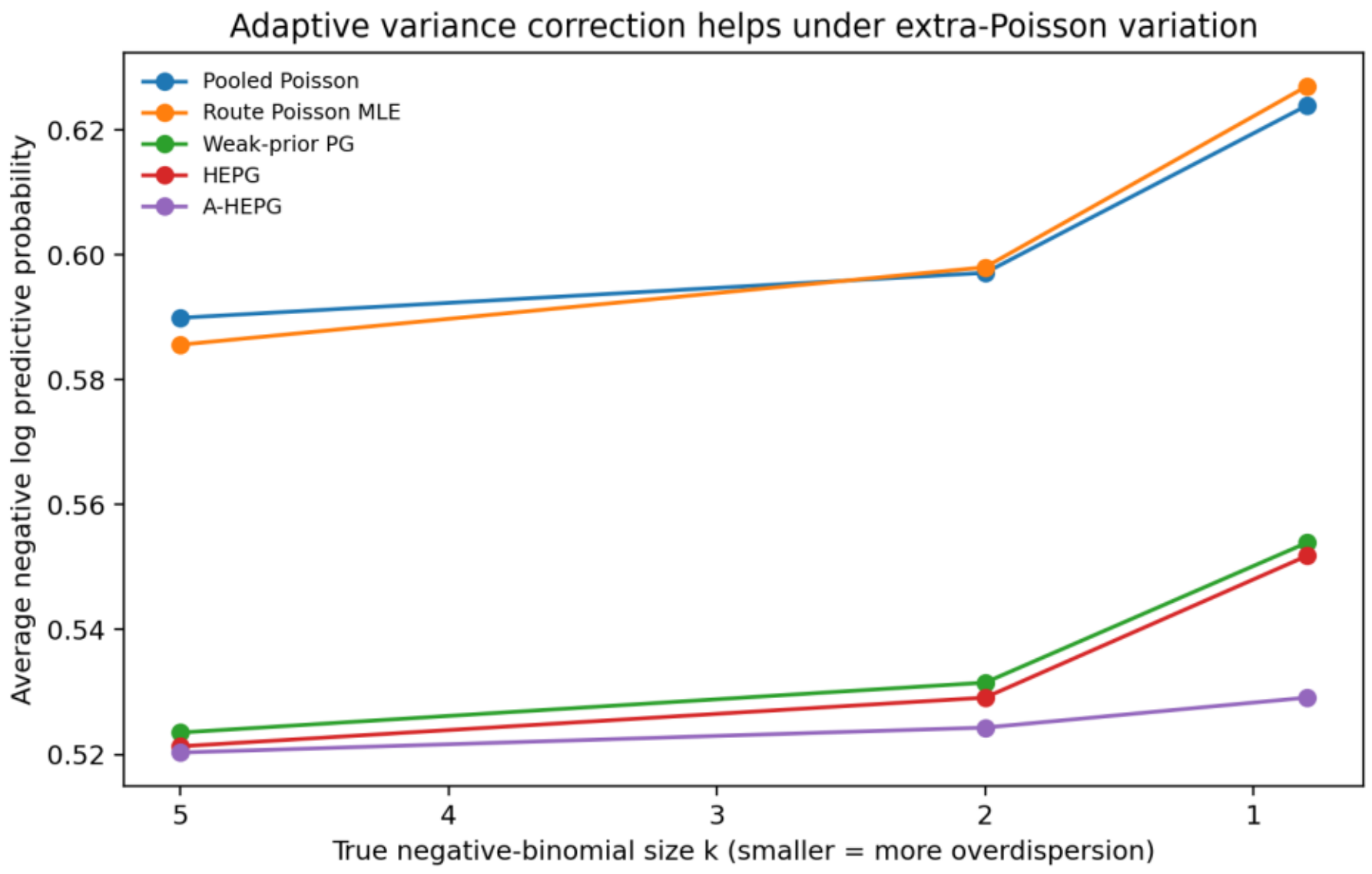

*Robustness to extra-Poisson variation. A-HEPG expands predictive variance when the data indicate overdispersion.*

### Interpretation of interval coverage

The central 90% predictive intervals cover more than 90% of outcomes in these very sparse experiments. The conservatism is largely a consequence of discreteness: when most predictive mass lies on 0 and 1, an equal-tailed integer interval cannot achieve every nominal coverage level exactly. For operational use, proper distributional scores and decision-specific quantiles are more informative than demanding exact nominal coverage from a discrete interval.

## Applied decision example: premium-cabin capacity sharing

To illustrate how a sparse predictive distribution can enter a downstream decision, consider a flight for which paid premium demand is forecast to be low while economy demand is strong. Some otherwise unused premium capacity can be made available to economy fare classes, with accepted economy customers accommodated in the premium cabin through complimentary upgrades near departure. Airline practice often calls this cabin floating; the academic literature commonly calls it premium-cabin capacity sharing (Escovar-Álvarez & Belobaba, 2022).

This example is useful statistically because the decision depends on the predictive tail, not only the predictive mean. The relevant quantity is the probability that an additional protected premium seat will ultimately be needed. HEPG and A-HEPG provide that probability directly through their full predictive distributions.

## A one-period seat-protection model

Let $C_P$ denote the number of premium seats available for the remaining booking horizon. Let $D_P$ denote remaining paid premium demand with predictive distribution obtained from HEPG or A-HEPG. Suppose economy demand is sufficiently high that every floated seat can be sold. Let $p_E$ be the net revenue from accepting one additional economy booking and $p_P$ be the net revenue from one premium booking. If a complimentary upgrade carries incremental variable cost $c_U$, replace $p_E$ below by $g_E = p_E - c_U$.

Choose $y \in \{0, \dots, C_P\}$ premium seats to protect for paid premium demand. Then

$$x = C_P - y$$

seats are floated to economy. Ignoring revenue from the fixed economy cabin because it is common to all choices, the expected relevant revenue is

$$V(y) = p_P \, E[\min(D_P, y)] + g_E(C_P - y).$$

The marginal value of protecting the $y$th premium seat rather than floating it is

$$V(y) - V(y-1) = p_P \, P(D_P \geq y) - g_E.$$

Therefore the optimal protection level is

$$y^\star = \max\left\{y \leq C_P : P(D_P \geq y) \geq \frac{g_E}{p_P}\right\}, \qquad x^\star = C_P - y^\star.$$

The cabin-floating rule above provides a direct answer to the price-difference question. A larger premium fare relative to the economy fare lowers the threshold $g_E/p_P$, causing more premium seats to be protected and fewer seats to be floated. When the price gap is small, the opportunity cost of using a premium seat for an economy passenger is also small, so the policy becomes more aggressive.

Under HEPG, $D_P$ is negative binomial from the HEPG predictive distribution. Under a plug-in Poisson forecast, the same rule holds with a Poisson survival probability. The decision framework is therefore distribution-agnostic once a coherent predictive distribution is available.

## Price-difference cases

To illustrate the rule, consider a synthetic flight with $C_P = 12$ premium seats remaining in the decision horizon and abundant excess economy demand. Set the economy fare contribution to $p_E = \$200$ and assume no additional variable upgrade cost. The sparse-demand forecast is an HEPG negative-binomial predictive distribution with mean remaining premium demand equal to 4 and shape parameter $A = 3$, corresponding to variance

$$4 + \frac{4^2}{3} = 9.33.$$

The relatively wide variance represents substantial uncertainty around a low premium-demand forecast.

The table below solves the optimization for five premium-fare cases.

*Synthetic cabin-floating decisions for different premium–economy fare differences. Economy fare contribution is $200, premium capacity is 12 seats, and the HEPG predictive mean premium demand is 4.*

| Premium fare | Fare gap | Fare ratio | Protect premium | Float to economy | Expected gain vs. no sharing |
|---:|---:|---:|---:|---:|---:|
| $220 | $20 | 1.10 | 1 | 11 | $1,532 |
| $300 | $100 | 1.50 | 2 | 10 | $1,326 |
| $400 | $200 | 2.00 | 3 | 9 | $1,154 |
| $600 | $400 | 3.00 | 5 | 7 | $937 |
| $1,000 | $800 | 5.00 | 6 | 6 | $688 |

The result is monotone and intuitive. When the premium fare is only 10% above economy, the model protects one premium seat and floats eleven. At a premium fare three times the economy fare, five seats are protected. At a five-to-one fare ratio, six seats are protected even though mean premium demand is only four; the additional protection reflects the asymmetric cost of spilling a high-value premium customer.

The “expected gain versus no sharing” column should be interpreted only within the synthetic one-period model. It is the incremental contribution from making otherwise unavailable premium capacity accessible to economy demand, net of the expected premium revenue displaced by the protection choice. It excludes the revenue of the base economy cabin and does not incorporate customer-experience effects, loyalty consequences, or reaccommodation costs.

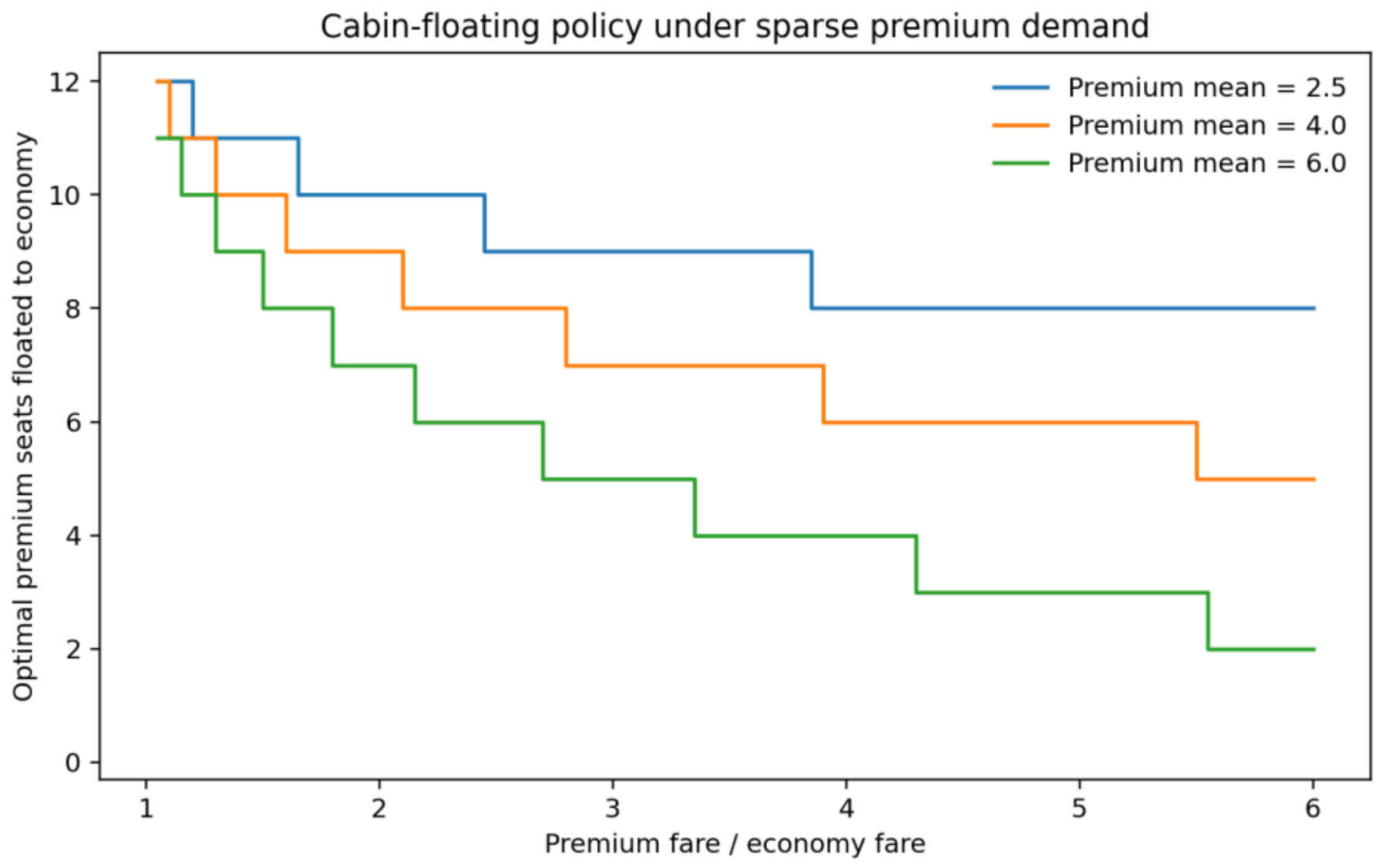

*Optimal number of premium seats floated to economy as the premium-to-economy fare ratio changes. Curves use an HEPG negative-binomial predictive distribution with posterior shape* $A = 3$ *and premium capacity of 12 seats. Higher expected premium demand or a larger premium fare gap causes more premium capacity to be protected.*

## Uncertain excess economy demand

If excess economy demand is not guaranteed, let $D_{E,\text{excess}}$ denote demand above the physical economy capacity. The marginal value of the $k$th floated seat becomes

$$g_E \, P\left(D_{E,\text{excess}} \geq k\right).$$

The protection comparison is then

$$p_P P(D_P \geq y) \quad \text{versus} \quad g_E P\left(D_{E,\text{excess}} \geq k\right).$$

Thus the same sparse-count framework can forecast both sides of the trade: low premium demand and uncertain overflow economy demand. The simple rule in the cabin-floating rule above is recovered when excess economy demand is effectively certain.

## Dynamic interpretation

The rule is most naturally reevaluated as departure approaches. Early in the booking horizon, uncertainty about high-value premium demand is larger and premium seats should generally be protected more conservatively. Late in the horizon, if the HEPG posterior continues to indicate weak premium demand while economy demand remains strong, more capacity can be released. This timing logic is consistent with prior premium-cabin capacity-sharing research, which finds that aggressive sharing too early can displace premium demand (Escovar-Álvarez & Belobaba, 2022).

A production implementation would update $D_P$ whenever meaningful bookings, cancellations, schedule changes, fare changes, or availability changes occur. The result is a dynamic protection level rather than a fixed number of floated seats for the entire booking horizon.

# Applied statistical interpretation

## Zero-history behavior and partial pooling

Consider a thin market with no observed requests in a short training history. Route-level Poisson maximum likelihood sets $\hat{\lambda} = 0$ and therefore assigns zero probability to every positive future outcome. Numerical clipping avoids infinite log loss in software but does not solve the conceptual problem.

HEPG treats "zero observed so far" as weak evidence for a low rate, not proof of a zero rate. The posterior estimate is

$$E(\lambda_i \mid Y_i = 0) = \frac{a}{b + S_i},$$

which decreases with zero-exposure history but remains positive. This behavior is natural for newly launched routes, infrequent services, thin fare products, or narrowly defined booking buckets.

## Forecast outputs

For applied use, the forecast should expose more than a point estimate. For each forecast cell, the system can return:

- posterior mean demand;
- $P(Y = 0)$ and $P(Y \geq 1)$;
- predictive quantiles such as the 50th, 80th, 90th, or 95th percentile;
- effective exposure $S_i$;
- shrinkage weight $w_i$ from the equation above;
- estimated overdispersion $k$; and
- a diagnostic flag for excess zeros or censoring.

The shrinkage weight is particularly interpretable in an applied analysis: it summarizes how much the unit's own history versus the population distribution contributes to the estimate, and therefore provides a useful measure of information strength.

## Diagnostic model selection in deployment

A practical implementation need not impose one count distribution on every cell. A diagnostically driven hierarchy is:

1. Use HEPG for sparse count cells with no strong residual overdispersion.
2. Use A-HEPG or a fitted negative-binomial hierarchy when extra-Poisson variance persists.
3. Use a hurdle or zero-inflated model only when zeros remain excessive after known zero-exposure and closed-availability records are handled.
4. For high-volume cells, compare the count models with mature airline forecasting methods; at sufficiently high counts the practical difference between discrete and continuous approximations may be small.

This diagnostic escalation strategy is preferable to selecting a highly flexible count model before examining whether the additional structure is supported by the data.

# Recommended validation with observational airline data

The present preprint is deliberately simulation-based. Before substantive operational conclusions are drawn, the framework should be evaluated on observational booking-request data using rolling-origin validation. An ideal dataset contains request counts, exposure or departure opportunities, availability status, booking horizon, market identifiers, calendar features, and censoring indicators.

Public U.S. Department of Transportation traffic data can illustrate cross-route heterogeneity and the prevalence of low-volume markets, but transported-passenger counts are not the same statistical object as latent booking demand. Consequently, such data should not be presented as unconstrained request counts without qualification. Stronger validation would use anonymized request or booking data after appropriate treatment of censoring and availability.

A recommended empirical protocol is:

1. Define the exact demand cell and forecast horizon before model fitting.
2. Remove zero-exposure periods and reconstruct availability.
3. Mark or unconstrain censored observations.
4. Split chronologically; do not randomly shuffle time.
5. Evaluate both point and probabilistic scores.
6. Stratify results by mean demand, zero fraction, route age, and effective exposure.

7. Report decision-sensitive metrics, such as quantile loss or revenue impact, only after the statistical forecast comparison is complete.

## Limitations

This study has several limitations. First, the reported numerical results are synthetic and therefore establish behavior under controlled data-generating mechanisms rather than performance in a production airline system. Second, the simulations use a relatively simple weekly seasonal structure and independent unit-level intensities; real data contain substitution, network effects, events, price responses, and nonstationarity. Third, A-HEPG is a moment-matched predictive variance correction rather than a fully specified hierarchical negative-binomial model. Fourth, structural zeros and censoring are discussed but not estimated in the simulation. Finally, a single shared prior may be inadequate in a highly heterogeneous network, motivating clustered or multilevel extensions.

These limitations define the scope of the current applied-statistics contribution and motivate direct empirical extensions.

## Extensions

Several extensions are natural from an applied-statistics perspective.

### *Dynamic route intensity.*

Replace a static $\lambda_i$ with a latent state $\lambda_{it}$ that evolves over time. Poisson–Gamma state-space models are especially attractive because they preserve much of the count interpretation and can support sequential updating.

### *Hierarchical market families.*

Let routes borrow strength first from a cluster—for example, region, haul length, hub status, or business/leisure composition—and then from the full network. This produces multilevel shrinkage instead of a single global Gamma prior.

### *Booking-curve structure.*

For days-to-departure forecasting, let $s_{it}$ include booking-horizon basis functions or cumulative arrival curves. The same route-level Gamma prior can then stabilize thin booking buckets.

### *Price and availability.*

Covariates can include fare, displayed availability, competitor schedule, or other demand drivers. Care is required because price and inventory are endogenous to expected demand.

*Decision-focused scoring.*

If the forecast is used for a particular protection or schedule decision, predictive distributions can be evaluated with asymmetric quantile loss or downstream revenue simulation in addition to generic proper scores.

## Conclusion

Sparse airline demand is best treated as a low-count inference problem rather than simply as a small-scale version of high-volume forecasting. A Poisson likelihood is a principled baseline because it respects nonnegative integer support and assigns probability to zero, but independent Poisson estimation is brittle when unit histories are short.

The HEPG framework stabilizes unit-level intensities through empirical-Bayes partial pooling and produces a closed-form negative-binomial posterior predictive distribution that propagates rate uncertainty. In the synthetic study, HEPG improves proper probabilistic scores relative to complete pooling and route-by-route Poisson maximum likelihood. When extra-Poisson variation is introduced, the A-HEPG correction improves robustness while leaving the correctly specified Poisson case nearly unchanged.

The premium-cabin capacity-sharing example illustrates why distributional forecasts can matter for decisions: the number of seats to release depends on predictive survival probabilities in the upper tail and on the relative cost of displacing high-value demand. The broader methodological message is not to use Poisson models indiscriminately, but to match the count likelihood to the support, pool information when individual histories are weak, propagate parameter uncertainty, distinguish true zeros from censoring or zero exposure, and move to negative-binomial, hurdle, or zero-inflated structure only when diagnostics justify the additional complexity. These principles extend beyond airline demand to sparse applied count problems more generally.

## Data and code availability

All numerical results in this preprint are generated from synthetic data. Supplementary information, reproducibility files, simulation summaries, example synthetic data, and figures are archived on Zenodo at [https://doi.org/10.5281/zenodo.23175101](https://doi.org/10.5281/zenodo.23175101) (DOI: 10.5281/zenodo.23175101). No proprietary airline data are used.

## Derivation of the HEPG predictive distribution

Given the posterior

$$\lambda_i \mid \mathcal{D} \sim \mathrm{Gamma}(A_i, B_i)$$

and a future count with exposure multiplier $c_{i,f}$,

$$Y_{i,f} \mid \lambda_i \sim \text{Poisson}(c_{i,f}\lambda_i),$$

integrating over the posterior uncertainty in $\lambda_i$ gives

$$P(Y_{i,f} = y \mid \mathcal{D}) = \frac{\Gamma(A_i + y)}{\Gamma(A_i)\, y!} \left(\frac{B_i}{B_i + c_{i,f}}\right)^{A_i} \left(\frac{c_{i,f}}{B_i + c_{i,f}}\right)^{y}.$$

This is the negative-binomial predictive distribution used by HEPG.

## Moment match for A-HEPG

Let posterior route intensity be

$$\lambda \sim \text{Gamma}(A, B)$$

and let independent multiplicative shock

$$U \sim \text{Gamma}(k, k).$$

For

$$Y \mid \lambda, U \sim \text{Poisson}(c\lambda U),$$

the mean is

$$m = \frac{cA}{B}.$$

Using independence,

$$E(\lambda^2) = \frac{A(A+1)}{B^2}, \qquad E(U^2) = \frac{k+1}{k}.$$

Therefore

$$\text{Var}(Y) = m + \frac{c^2 A(k + A + 1)}{B^2 k}.$$

A negative-binomial distribution with mean $m$ and size $r$ has variance $m + m^2/r$. Equating the extra-variance terms gives

$$r = \frac{Ak}{k + A + 1}.$$

## Reproducibility notes

The supplied Python script uses fixed random seeds, 400 synthetic markets per replication, 30 replications per scenario, and the parameter settings described in the simulation study section. The summary CSV files included with the source package correspond to those runs.

Small numerical differences can occur across library versions because the empirical-Bayes hyperparameters are obtained by numerical optimization.